\documentclass[aps,prb,preprint,superscriptaddress]{revtex4-2}
\usepackage{graphicx}  
\usepackage{dcolumn}   
\usepackage{bm}        
\usepackage{amssymb}
\usepackage[english]{babel}
\usepackage{booktabs}
\usepackage{array}
\usepackage{caption}
\usepackage{subcaption}
\usepackage{hyperref}
\hypersetup{
	colorlinks=true,
	linkcolor=green,
	filecolor=magenta,      
	urlcolor=blue,}
\begin{document}
\title{Structural transition and emission enhancement in vacancy ordered halide double perovskite Cs$_2$TeCl$_6$ under pressure}

\author{Suvashree Mukherjee}
\affiliation{Department of Physical Sciences, Indian Institute of Science Education and Research Kolkata, Mohanpur Campus, Mohanpur 741246, Nadia, West Bengal, India.}
\affiliation{National Centre for High Pressure Studies, Indian Institute of Science Education and Research Kolkata, Mohanpur Campus, Mohanpur 741246, Nadia, West Bengal, India.}

\author{Debabrata Samanta}
\email [Present address:]{Beijing Academy of Quantum Information Sciences, Beijing, 100193, P.R. China}
\affiliation{Department of Physical Sciences, Indian Institute of Science Education and Research Kolkata, Mohanpur Campus, Mohanpur 741246, Nadia, West Bengal, India.}
\affiliation{National Centre for High Pressure Studies, Indian Institute of Science Education and Research Kolkata, Mohanpur Campus, Mohanpur 741246, Nadia, West Bengal, India.}

\author{Bidisha Mukherjee}
\affiliation{Department of Physical Sciences, Indian Institute of Science Education and Research Kolkata, Mohanpur Campus, Mohanpur 741246, Nadia, West Bengal, India.}
\affiliation{National Centre for High Pressure Studies, Indian Institute of Science Education and Research Kolkata, Mohanpur Campus, Mohanpur 741246, Nadia, West Bengal, India.}

\author{Konstantin Glazyrin}
\affiliation{Photon Science, Deutsches Elektronen Synchrotron, 22607 Hamburg, Germany}

\author{Goutam Dev Mukherjee}
\email [Corresponding author:]{goutamdev@iiserkol.ac.in}
\affiliation{Department of Physical Sciences, Indian Institute of Science Education and Research Kolkata, Mohanpur Campus, Mohanpur 741246, Nadia, West Bengal, India.}
\affiliation{National Centre for High Pressure Studies, Indian Institute of Science Education and Research Kolkata, Mohanpur Campus, Mohanpur 741246, Nadia, West Bengal, India.}
\date{\today}

\begin{abstract} 
The effect of pressure on the structural evolution, enhancement of photoluminescence intensity and optical band gap of a vacancy ordered double halide perovskite Cs$_2$TeCl$_6$ is investigated systematically. We use synchrotron x-ray diffraction, Raman spectroscopy, optical band gap and photoluminescence measurements to explore the structural and the optical properties of Cs$_2$TeCl$_6$ under pressure up to 30.0 GPa. We find that Cs$_2$TeCl$_6$ undergoes a structural transition from cubic Fm$\bar{3}$m to monoclinic P2$_1$/n at very low pressure below 1.0 GPa. A significant increase in photoluminescence intensity and a rapid decrease in optical band gap are observed, which are related to the octahedral distortion and structural transition. Interestingly, with increasing pressure, the sample regains its ambient Fm$\bar{3}$m structure at around 3.4 GPa maintaining the cubic phase up to 30.0 GPa. The sample undergoes an iso-structural transition at around 14.1 GPa pressure with a slight decrease in compressibility.
	
\end{abstract}

\maketitle

\section{INTRODUCTION}
The lead halide perovskites demonstrate good performance in various optoelectronic applications including light-emitting diodes, lasers, and photovoltaics due to their exceptional properties. In spite of these advantages, poor stability and the toxicity of lead limit its practical applications. Therefore, lead-free stable halide perovskites for suitable optoelectronic applications are in the focus of new research \cite{Jena,Pecunia}. In this regard, lead-free halide double perovskite with the formula unit A$_2$BB$^\prime$X$_6$ (where A is an alkali or alkaline metal, B and B$^\prime$ are monovalent and trivalent cations, respectively and X is halogen) have attracted the attention of the scientific community due to their satisfactory performance in the optoelectronic field \cite{Ghosh,Muscarella}.

The vacancy-ordered halide double perovskites (VOHDP) are a defect-ordered variant of the perovskite structure, where B site is replaced by tetravalent cation and B$^\prime$ site is vacant resulting in isolated BX$_6$ octahedra bound electrostatically by A cations. The absence of polyhedral connectivity in VOHDPs leads to lattice anharmonicity, which creates strong electron-phonon coupling \cite{Preeti}. The optical absorption and band gap are determined by electronic states of B and X site ions. Therefore, the VOHDPs can be tuned by suitable compositional modification at B and X sites to elicit desirable optical and electronic properties \cite{Maughan}. Several studies have been performed by replacing the B site with Sn$^{4+}$, Ti$^{4+}$ and Te$^{4^+}$ cations \cite{Muscarella,Preeti,Maughan}. In Te-based VOHDPs, Te$^{4+}$ contains a 5s$^2$ lone pair which is suspected to play an important role in several optoelectronic processes including nonlinear harmonic generation, enhanced emission etc \cite{Brumberg}.

The Te-based VOHDP Cs$_2$TeCl$_6$ is a stable compound at ambient conditions and might be a potential candidate in nonlinear optoelectronics. However the indirect nature of its optical band gap with a large value ($\approx$ 2.54 eV) affects its performance as a photovoltaic material \cite{Jiang,Berri}. Because of the soft lattice of halide double perovskites, the application of external pressure could greatly impact the crystal structure of the material, allowing tuning of the optical band gap and improve its performance as a photovoltaic material. For example, Cs$_2$SnI$_6$ undergoes a structural transition from face-centred cubic (FCC) to monoclinic I2/m phase at around 8-10 GPa, whereas Cs$_2$SnCl$_6$ and Cs$_2$SnBr$_6$ retain the parent FCC structure under pressure \cite{Bounos}. The optoelectronic performance of Cs$_2$PtBr$_6$ can be enhanced under pressure with an optimal band gap of 1.34 eV at 12 GPa maintaining its structural stability \cite{Cao}. The indirect to direct band gap transition and increase of absorption cross-section is also possible under high pressure \cite{Zhao,Diwen}.

Therefore, it is important to investigate the pressure-induced evolution of structural and optical properties of Cs$_2$TeCl$_6$ before any practical applications. Shi et al.\cite{Shi} has synthesized Cs$_2$TeCl$_6$ using a metal halide precursor TeCl$_4$ and reported pressure-modulated self-trapped exciton (STE) emission but no structural transition up to 11 GPa. In the present paper we have synthesized the Cs$_2$TeCl$_6$ using a metal oxide precursor TeO$_2$. Recently, Brumberg et al. \cite{Brumberg} have reported the role of distinct precursor during the synthesis process of a similar type of VODHP Cs$_2$TeBr$_6$ on optical absorption edge at ambient conditions. The behaviour of Cs$_2$TeCl$_6$, synthesized using metal oxide precursor, under pressure has not been reported yet. Herein we have performed a systematic investigation of Cs$_2$TeCl$_6$ up to 30.0 GPa by combining powder x-ray diffraction (XRD), Raman scattering, optical band gap, and photoluminescence (PL) measurements. Our investigation shows that the Cs$_2$TeCl$_6$ undergoes two structural transitions at around 1.0 GPa and 3.4 GPa followed by an iso-structural transition at around 14.1 GPa. An enhancement of PL intensity is noticed at 1.2 GPa, which is related to the octahedral distortion. A rapid decrease in optical band gap by 10.7\% up to 2.9 GPa is observed.

\section{Experimental section}
Crystalline powders of Cs$_2$TeCl$_6$ are prepared by the acid precipitation method. CsCl (Purity $\geq$ 99.9\%) and TeO$_2$ (Purity $\geq$ 99.9 \%) are purchased from Sigma Aldrich. A mixture of CsCl and TeO$_2$ with a 2:1 molar ratio is dissolved in 2 ml HCl (48 wt \% in water) in a capped vial. The mixture is then continuously stirred for 2 hours at a temperature 110$^{\circ}$C. After that, the mixture is kept at the same temperature for 2 hours without stirring. Then the obtained solution is allowed to cool to room temperature and kept overnight. The precipitated compound Cs$_2$TeCl$_6$ is washed with ethanol several times and dried for experimental use.

Pressure-dependent Raman spectra, PL and optical band gap measurements are carried out by using a piston-cylinder type diamond anvil cell (DAC) with a culet of diameter 300 $\mu m$. A steel gasket of thickness 290 $\mu m$ is preindented to the thickness of about 50 $\mu m$ followed by drilling a 100 $\mu m$ hole at the centre of the indented portion to serve as the sample chamber. The sample along with a few grains of ruby are loaded in the sample chamber. Silicon oil is used as a pressure transmitting medium (PTM) and ruby fluorescence technique \cite{ruby} is employed for pressure calibration within the sample chamber.

Raman scattering measurements are performed using a confocal micro-Raman spectrometer (Monovista from SI GmBH) in the backscattering geometry with 750 mm monochromator and a back-illuminated PIXIS 100BR (1340$\times$100) charge-coupled device camera. The sample is excited with a Cobolt-samba diode pump laser light of wavelength 532 nm and the spectra are collected using a Bragg filter with grating 1500 grooves/mm. A 20X infinity-corrected long working distance objective is used to focus the incident radiation beam and also to collect the scattered radiations.

PL measurements are performed in backscattering geometry using Horiba Jobin-Yvon LabRAM HR-800 spectrometer with an 800 mm focal length achromatic flat field monochromator. The sample is excited with an Ar$^+$ ion laser of wavelength 488 nm and the spectra are collected with grating 600 grooves/mm. 

High pressure absorption measurements are performed using a custom-made Sciencetech 15189RD absorption spectrometer. To focus the broadband light on the sample, an achromatic lens with a focal length of 19 mm is used. The transmitted light is collected using a 10X objective. The absorbance is calculated as: $A= -\log(\frac { I_t -I_d}{I_0- I_d})$, where $I_0$ is the intensity of input light, $I_t$ is the intensity of light transmitted through the sample, and $I_d$ is the intensity at the dark environment. The bandgap is determined from the absorption edge using a Touc plot for indirect bandgap. The linear absorption edge is fitted by the equation: $(Ah\nu)^{0.5} = C \times (h\nu - E_g)$; where A is the absorbance, $h\nu$ is the energy of the photon, C is a constant, and $E_g$ is the indirect optical bandgap.  

High-pressure XRD measurements are carried out at the P02.2 beamline at PETRA III, Germany using a monochromatic x-ray beam of wavelength 0.2907 $\AA$ and spot size of $(8\times3) \mu m^2$. XRD images are collected by the 2D large area detector from Perkin Elmer (XRD1621). CeO$_2$ is used as a standard sample for the sample to the detector distance calibration. We have used a membrane-driven symmetric DAC and a rhenium gasket in that case. We have employed DIOPTAS \cite{dioptas} software for the conversion of 2D diffraction images to intensity versus 2$\theta$ plot. All the XRD data are analyzed by GSAS \cite{GSAS,EOSfit,Vesta} software.

\section{RESULTS AND DISCUSSION}

The Cs$_2$TeCl$_6$  powder is successfully synthesized in pure phase as confirmed by energy-dispersive x-ray spectroscopy (EDX) and synchrotron XRD measurements. The EDX spectra with atomic percentage is shown in the Supplementary FIG.S1. The ambient XRD pattern could be indexed to a cubic structure and the Rietveld refinement resulted in a good fit to space group Fm$\bar{3}$m with lattice parameters a=b=c=10.4016(3)$\AA$, which agrees well with previously reported literature \cite{chaojiepi}. In the unit cell, Cs and Te atoms occupy the high symmetry 8c and 4a Wyckoff positions with coordinates (0.25, 0.25, 0.25) and (0,0,0) while Cl atoms are situated at 24e Wyckoff positions with coordinates (x,0,0) with x being a free parameter. Each Te atom is surrounded by Cl atoms in the octahedra and the Cs atom resides in the hollow between TeCl$_6$ octahedra, with 12-fold coordination of the Cl atom in ambient conditions.

The pressure evolution of the XRD pattern at selected pressures	is shown in FIG.2. As pressure increases, new peaks appear at 2$\theta$ =4.8$^{\circ}$ and 8.4$^{\circ}$ at 1.0 GPa, which merge with background after 2.6 GPa. The XRD pattern is reindexed and a good fit to a monoclinic lattice in space group P2$_1$/n with lattice constants a=7.255(5)$\AA$, b= 7.351(5)$\AA$, c=10.300(5)$\AA$ and $\beta$=90.01(9)$^{\circ}$ are obtained. A similar VODHP K$_2$SnCl$_6$ is known to undergo cubic to monoclinic structural transition at low temperature (190K) \cite{boysen}. We have carried out Rietveld refinement of the XRD pattern of Cs$_2$TeCl$_6$ at 1.0 GPa using starting atom position parameters given in the Boysen et al. \cite{boysen}. The final refined atom positions are listed in Table I. It can be seen that this phase is formed due to the tilting of isolated TeCl$_6$ octahedra about the a-axis as shown in the FIG.1(d). The monoclinic structure persists up to 2.6 GPa. Interestingly, above 3.4 GPa, we find that the sample comes back to the original cubic Fm$\bar{3}$m phase. Therefore it looks like the monoclinic structure is a transient phase, which is introduced due to strain in the lattice.  We do not find any other structural transition till 30.5 GPa, the highest pressure of this study. The lattice volume continuously decreases with applying pressure above 3.4 GPa in the cubic Fm$\bar{3}$m structure as shown in FIG.3(a). Interestingly, we find that there is a change in slope in the pressure behaviour of volume at about 14.1 GPa. The volume vs pressure data for the whole region resulted in a poor fit using a single Birch-Murnaghan equation-of-state (EoS)(Supplementary FIG.S3). To investigate the effect of internal strain we have plotted reduced pressure $H=\frac{P}{3f_E(1+2f_E)^\frac{5}{2}}$ with respect to the Eulerian strain $f_E= \frac{1}{2}[(\frac{V_0}{V})^\frac{2}{3}-1]$. The $H$ vs $f_E$ plot indicates the internal strain within the lattice increases abruptly at around that pressure point 14.1 GPa. Therefore, two sets of 3rd-order Birch-Murnaghan EoS are fitted to the volume vs pressure data, one set in the pressure range of 3.4 GPa to 14.1 GPa and the second set above 14.1 GPa. The EoS fit results slight increase in bulk modulus (B$_0$) and its derivate (B$^\prime$). The obtained values of B$_0$ and B$^\prime$ are 30(3) GPa and 2.5(6) respectively below 14.1 GPa whereas those values are 34.9(4) GPa and 4.07(6) above 14.1 GPa. Such a decrease in compressibility without a structural phase transition is very interesting and may be due to large internal lattice strain. This is reflected in the observed XRD patterns, which broaden with increasing external pressure. Very broad XRD pattern at 30.5 GPa indicates a large disorder being induced due to external pressure.

As a complementary study, we have performed high-pressure Raman spectroscopic measurements.The Cs$_2$TeCl$_6$ shows four Raman active modes P$_1$ (T$_{2g}$), P$_2$ (T$_{2g}$), P$_3$ (E$_g$) and P$_4$ (A$_{1g}$) at  47.2, 137.3, 242.6 and 286.0 cm$^{-1}$, respectively at ambient conditions. The P$_1$ mode is assigned to the translational motion of Cs$^+$ ion. The origin of  P$_2$ mode is related to the bending and P$_3$, P$_4$ modes are related to asymmetric and symmetric stretching of the TeCl$_6$ octahedra \cite{Jiang}. The pressure evolution of Raman spectra at some selected pressure values is shown in FIG.4(a). At 0.1 GPa, noticeable changes are observed. P$_1$ mode is not detectable and a new mode P$_5$ appears at 0.1 GPa. The new mode P$_5$ persists up to 2.5 GPa and is not detectable above 3.6 GPa. Interestingly, P$_1$ mode reappears at 3.6 GPa and its intensity increases continuously up to 16.1 GPa, after which, the intensity of P$_1$ mode decreases. All the Raman modes exhibit blue shift upon compression and the modes start broadening at about 20.0 GPa and merge with the background above 25.4 GPa. The disappearance and reappearance of P$_1$ mode is possibly related to the cubic to monoclinic structural transition as suggested by XRD analysis. Slight softening and disappearance of P$_5$ mode are also an indication of structural transition. Pressure evolution of the Raman shift shows a clear slope change at around 2.5 GPa in the linear pressure behavior. The slopes of P$_2$ and P$_4$ mode decrease whereas that of P$_3$ mode increases. The results might be attributed to the structural transition from the monoclinic to the cubic phase. It is also noticeable from FIG.4(b) that all the Raman modes show a slope change in their linear variation with pressure at 14.1 GPa. The slope of P$_2$ mode is 6.16(8) cm$^{-1}$/ GPa below 14.1 GPa and drops to 3.67(12) cm$^{-1}$/ GPa above 14.1 GPa. The slopes of P$_1$, P$_3$ and P$_4$ mode also decrease above 14.1 GPa, which is an indication of a decrease in compressibility above this pressure point. The continuous blue shift of all Raman modes (P$_1$, P$_2$, P$_3$ and P$_4$) with pressure indicates a continuous decrease in interatomic distances accompanied by a decrease in unit cell volume. The result is compatible with the XRD data. The Raman spectrum taken after pressure release does not match with the ambient pattern obtained before increasing the pressure, rather it looks quite similar with the Raman pattern at 0.1 GPa. Hence, the material does not seem to regain its original phase completely after removing the pressure. 

The UV-visible absorption spectra at ambient conditions and some selected pressure values are shown in FIG.5(a). The absorption edge shows a red shift up to 2.9 GPa, after which the absorption edge shows a slight blue shift up to 4.8 GPa followed by a continuous redshift. The ambient optical band gap of Cs$_2$TeCl$_6$, determined from the absorption spectra is 2.51(4) eV, which is closer to the previously reported value \cite{chaojiepi}. The variation of the optical band gap with pressure is shown in FIG.5(b). The band gap initially reduces under compression up to 2.9 GPa, after which it increases with pressure up to 4.8 GPa followed by a continuous reduction up to 32.0 GPa. The valence band maximum (VBM) and conduction band minimum (CBM) of the Cs$_2$TeCl$_6$ are mainly dominated by the electronic states of TeCl$_6$ octahedra at ambient condition. The VBM is mainly composed of the Cl 3p orbital and a small amount of the Te 5s orbital, while the CBM is dominated by Te 5p and Cl 3p orbital \cite{Jiang}. Therefore, the Te-Cl bond length seems to play a crucial role in the variation of this band gap. The variation of the average bond length of TeCl$_6$ octahedra with pressure is shown in FIG.5(b), which looks quite similar to the variation of the optical band gap under pressure. The Te-Cl bond length increases in the pressure range of 2.9 GPa to 4.8 GPa and decreases after that. The decrease in Te-Cl bond length causes orbital overlap resulting in a decrease of band gap above the pressure  2.9 GPa. The decrease in optical band gap value from ambient to 2.9 GPa might be associated with the structural transition from cubic to monoclinic structure. After 2.9 GPa, the Cs$_2$TeCl$_6$ crystal consisting of undistorted TeCl$_6$ octahedra mainly depends on the Te-Cl bond length.

Cs$_2$TeCl$_6$ exhibits a broad emission band at 599 nm with a full-width half maximum (FWHM) of 123 nm accompanied by a large Stokes shift 0.8 eV at ambient conditions, which correspond to the self-trapped excitonic (STE) emission \cite{Jiang}. The pressure-dependent PL spectra at some selected pressure points are shown in the FIG.6(a) and (b). The PL intensity increases with pressure up to 1.2 GPa followed by a continuous decrease up to 16.2 GPa, after which it vanishes. To investigate emission behaviour under pressure, PL peak position, integrated intensity, and FWHM are estimated by fitting each PL spectrum to a Gaussian function. The PL intensity at 1.2 GPa pressure is almost 6 times that at ambient pressure. In halide perovskites, emission enhancement under pressure has mostly been explained by the octahedral distortion \cite{plreview}. We have calculated the distortion index D, from bond lengths, which is defined as $D=\frac{1}{n}\sum_{i=1}^{n}\frac{l_i-l_a}{l_a}$, where $l_i$ is the distance from the central atom to the ith coordinating atom, and $l_a$ is the average bond length \cite{distortion}. The distortion indices are 0.082 (3) at 1.0 GPa and 0.081(3) at 2.6 GPa. The distortion index becomes 0 at 3.4 GPa since the Cs$_2$TeCl$_6$ regains the cubic structure. The enhancement of PL intensity in this case also may be attributed to the distortion of TeCl$_6$ octahedra since XRD analysis suggests the crystal goes through maximum octahedral distortion at around 1.0 GPa. The PL peak position shows a continuous blue shift with increasing pressure, which indicates the energy level of STE state rises continuously. The FWHM of PL decreases under compression, indicating continuous decrease in the strength of electron-phonon coupling under pressure \cite{Iaru}. Our results differ from that mentioned by Shi et al.\cite{Shi} in the following aspects. No new peak is reported in pressure dependent XRD and Raman spectra and they have predicted transitions of electronic states at 1.6 GPa and 5.8 GPa. In contrast, current results show clear indications of structural transitions and enhancement of PL due to increased distortion in the unit cell. Hence, the response of Cs$_2$TeCl$_6$ under pressure seems to depend on the reactants used during the synthesis procedure.

\section{Conclusions}
Cs$_2$TeCl$_6$, consisting of isolated TeCl$_6$ octahedra, undergoes a structural transition from cubic (Fm$\bar{3}$m) to monoclinic (P2$_1$/n) phase with distorted octahedra below 1.0 GPa. The octahedral distortion results in a significant increase in PL intensity at around 1.2 GPa and a rapid reduction of the optical band gap. The octahedral distortion lowers with increasing pressure since the Cs$_2$TeCl$_6$ regains its cubic (Fm$\bar{3}$m) phase above 2.5 GPa. The Cs$_2$TeCl$_6$ also undergoes an iso-structural transition at around 14.1 GPa with a slight decrease in compressibility. Our results improve the understanding of the structural and optical properties of  Cs$_2$TeCl$_6$, which is anticipated to be beneficial for the development of optoelectronic devices using this type of material.

\section{Acknowledgments}
The authors acknowledge the financial support from the Department of Science and Technology, Government of India under India@DESY collaboration to carry out high-pressure XRD measurement at the P02.2 beamline of PETRA III, DESY, Germany. The financial support from the Science and Engineering Research Board (SERB), Government of India, Grant No. CRG/2021/004343 for the development of the lab based high-pressure UV-VIS-NIR Absorption Spectrometer is gratefully acknowledged. SM also acknowledges the fellowship grant provided by the CSIR, Government of India.

\begin{table}[h]
	\caption*{Table I}{Relative atomic positions at 1.0 GPa after Rietveld refinement} 
	
	\begin{tabular}{c@{\hskip 0.5in} c@{\hskip 0.5in} c@{\hskip 0.5in} c@{\hskip 0.5in} c@{\hskip 0.5in}}
		\hline\hline 
		Atom&Wyckoff position&x/a&y/b&z/c  \\ 
				
		\hline 
		Cs & 4e & 0.51269 (2) & 0.04795(4)  & 0.23707(2)\\
		
		 Te & 2a&0 & 0 & 0\\
		
		 Cl(1) & 4e& 0.23704(3) &  0.20155(4) &  0.00582(5)  \\
		
		 Cl(2)& 4e & 0.2443(5) &  -0.25722(4)& 0.03977(2)\\
		
		 Cl(3)	& 4e&  0.0964(1) & 0.04701(3) &0.26875(2)  \\

		\hline\hline
	\end{tabular}
\end{table}

\newpage

\begin{figure}[ht]
	\centering
	\includegraphics[scale = 0.6]{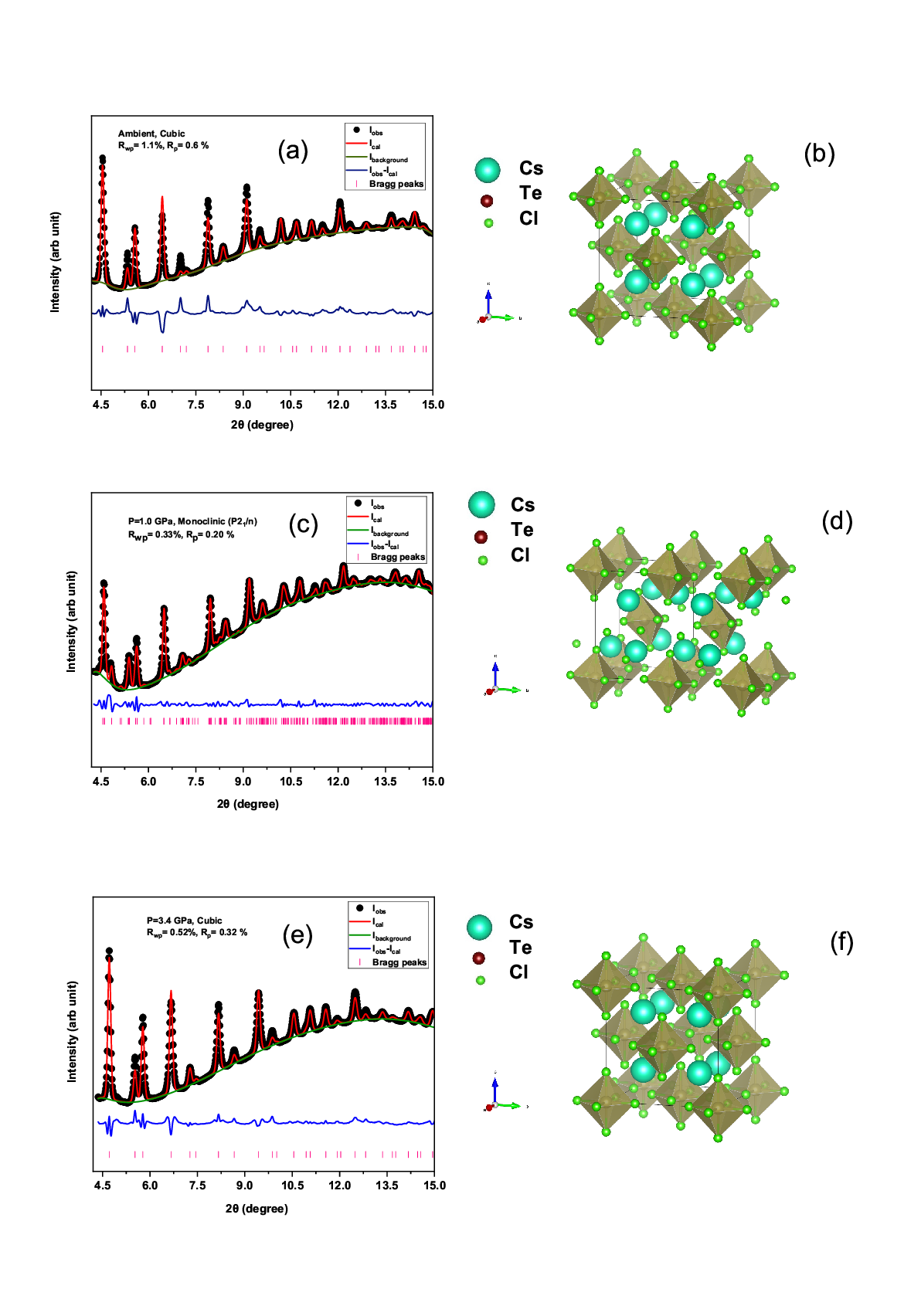}
	\caption{Rietveld refinement of the XRD pattern of Cs$_2$TeCl$_6$ at (a) ambient in cubic (Fm$\bar{3}$m),(c) 1.0 GPa in monoclinic (P2$_1$/n), and (e) 3.4 GPa in cubic (Fm$\bar{3}$m) structure. The black balls represent experimental data. Red, green, and navy lines are Rietveld fit to the experimental data, background, and difference between experimental and calculated data, respectively. The magenta vertical lines show the Bragg peaks of the sample. Schematic representations of the unit cell at (b) ambient conditions, (d) 1.0 GPa, and (f) 3.4 GPa are shown on the right side.}
\end{figure}

\begin{figure}[ht]
	\centering
	\includegraphics{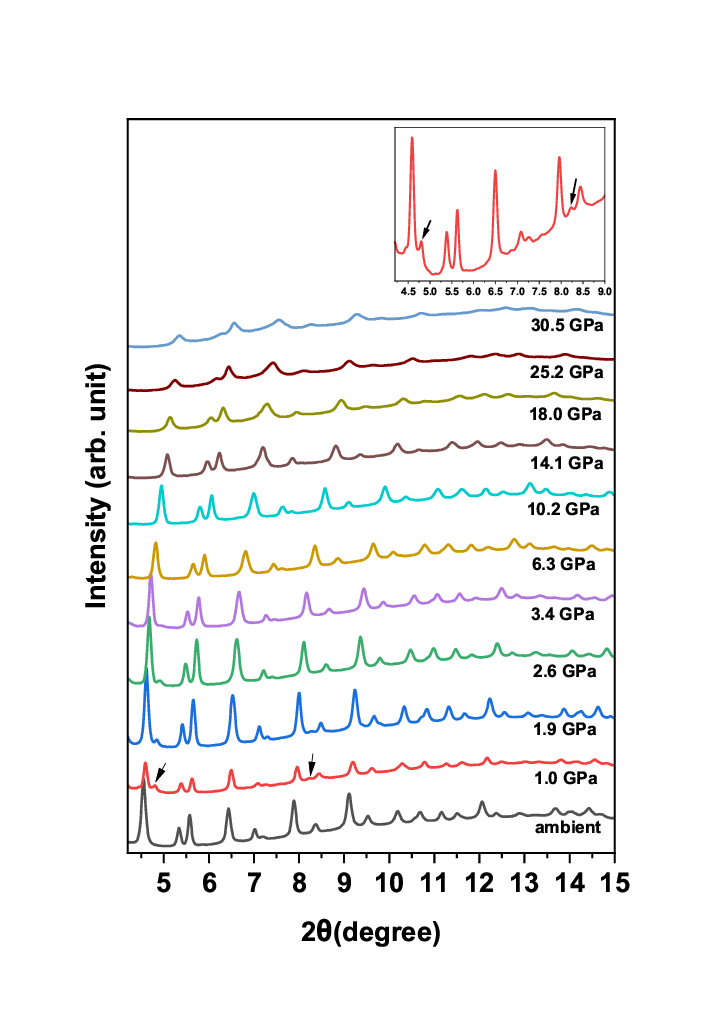}
	\caption{XRD pattern of Cs$_2$TeCl$_6$ at some selected pressures. Pressures (in GPa) are indicated alongside the respective data. The black arrows indicate the appearance of new peaks. The XRD pattern at 1.0 GPa is magnified in the right side at the top of the figure.}
\end{figure}

\begin{figure}[ht]
	\centering
	\includegraphics[scale=0.8]{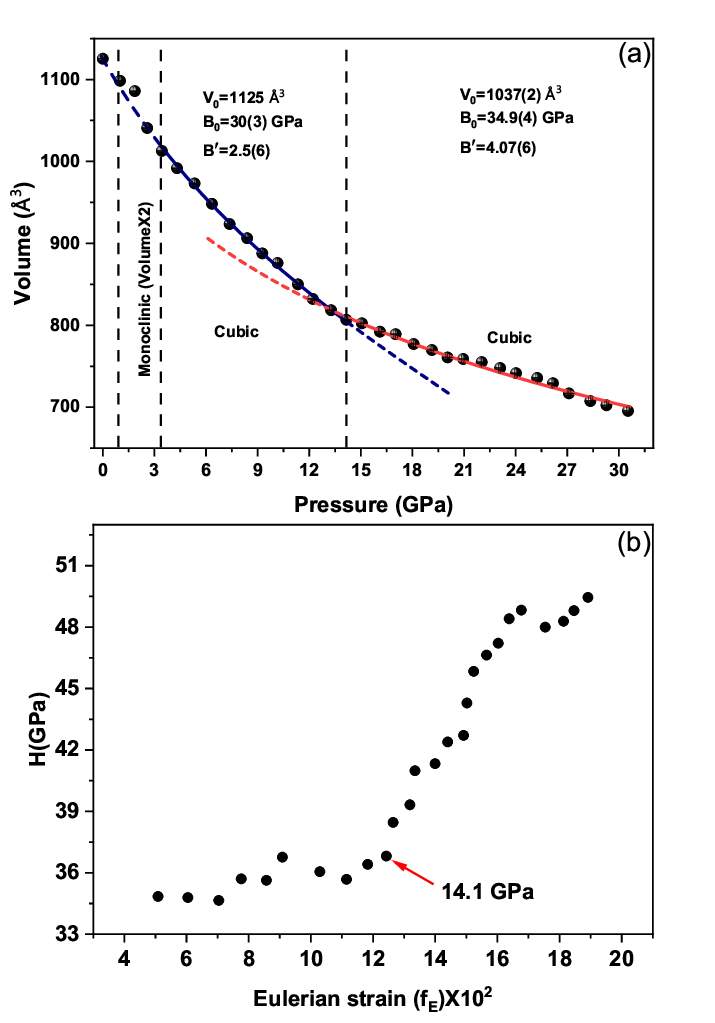}
	\caption{(a). Pressure evolution of the unit cell volume of Cs$_2$TeCl$_6$. The lines passing through the volume is the 3rd order Birch–Murnaghan EoS fit to the volume data. (b) Plot of normalized pressure (H) vs. Eulerian strain (f$_E$)}
\end{figure}

\begin{figure}[ht]
	\centering
	\includegraphics[scale=0.8]{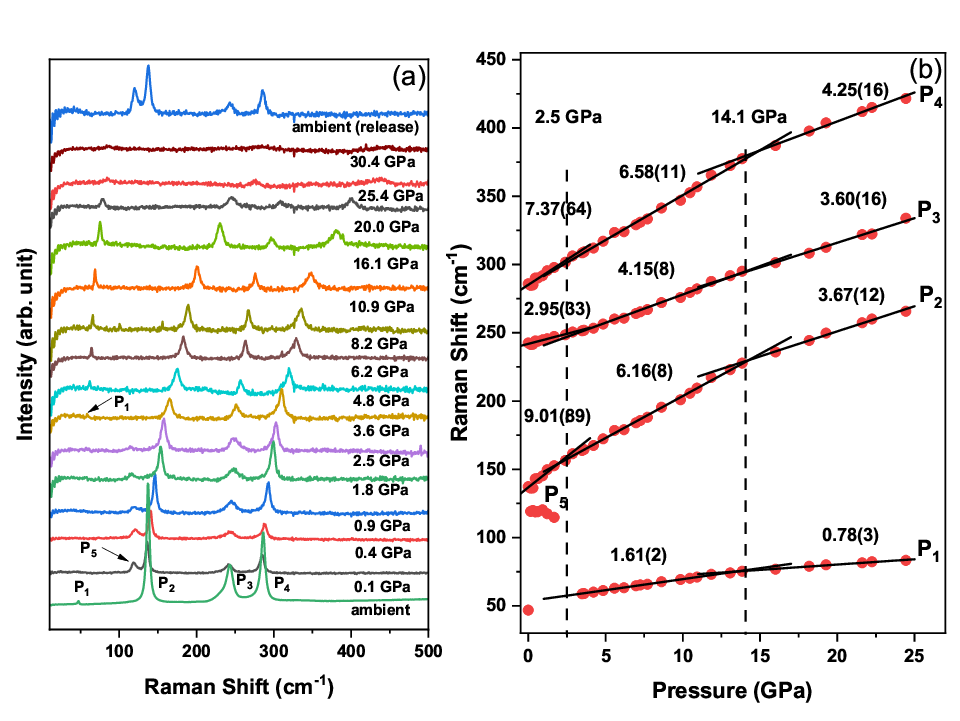}
	\caption{(a). Raman spectra of Cs$_2$TeCl$_6$ at selected pressure points. The new mode P$_5$ is marked by black arrow. (b) Pressure evolution of Raman shift. Lines passing through the data points are the linear fit to the data.}
\end{figure}

\begin{figure}[ht]
	\centering
	\includegraphics[scale=0.8]{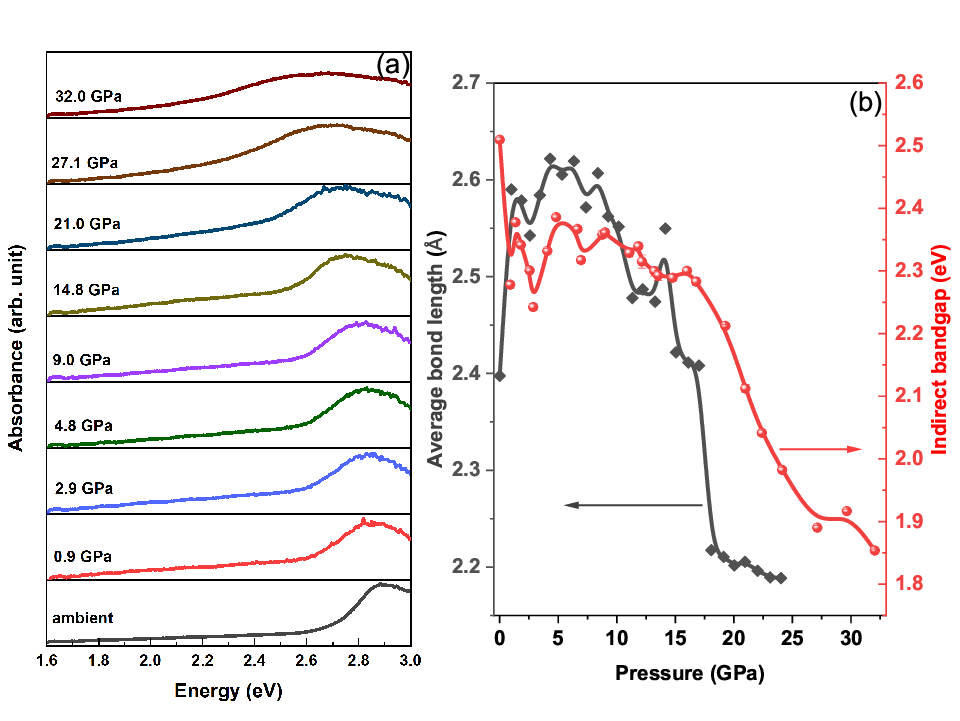}
	\caption{(a). UV-Vis absorption spectra of Cs$_2$TeCl$_6$ at some selected pressures.(b) Variation of optical bandgap under pressure.}
\end{figure}

\begin{figure}[ht]
	\centering
	\includegraphics[scale=0.8]{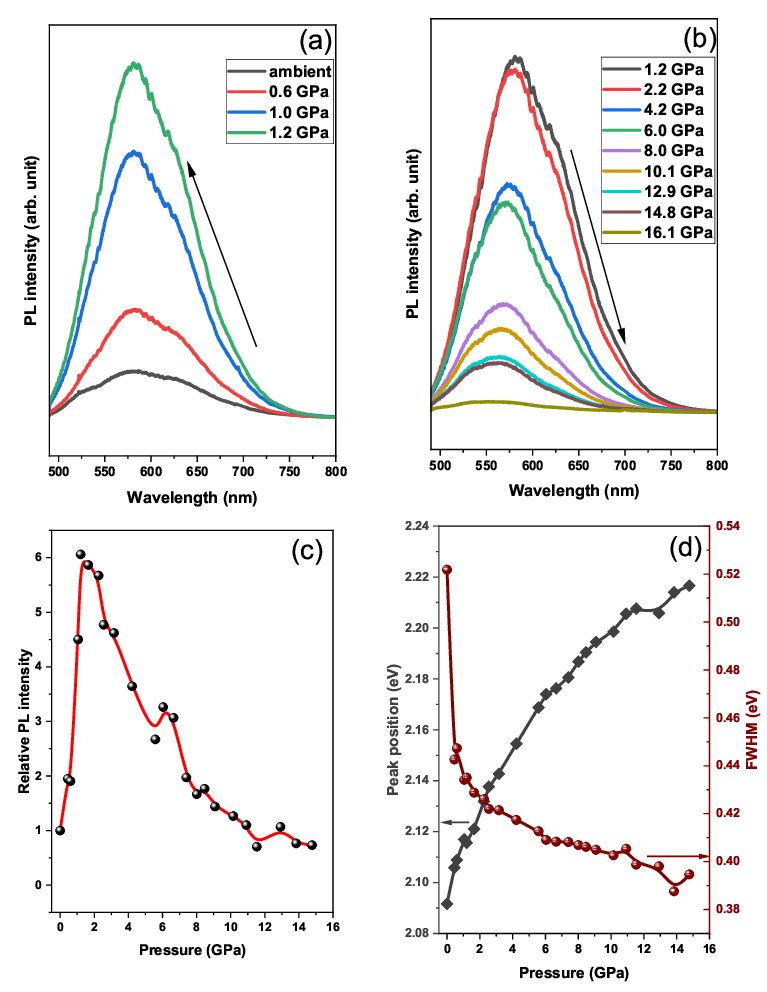}
	\caption{(a) and (b) Pressure-dependent PL spectra of Cs$_2$TeCl$_6$ (c) Relative intensity of PL at different pressures. (d) PL peak position and FWHM of PL spectra as a function of pressure.}   
\end{figure}

\begin{figure}[ht]
	\centering
	\includegraphics[scale=0.8]{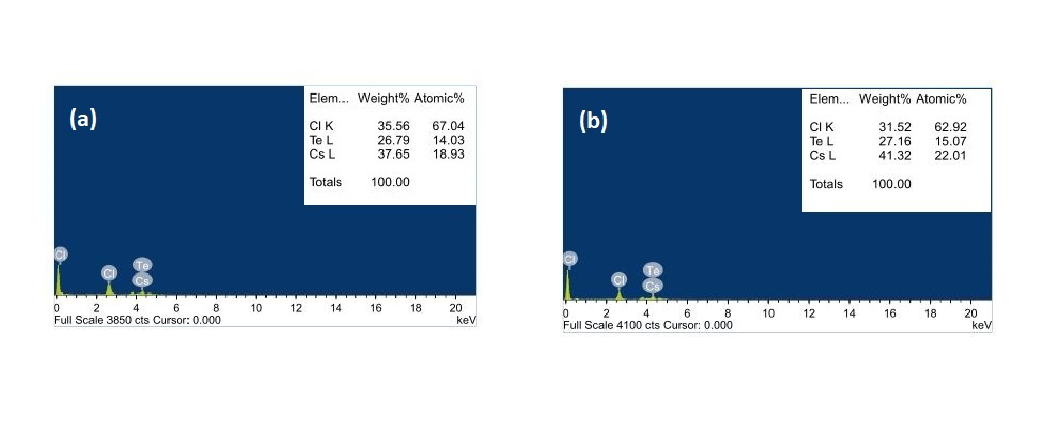}
	\vspace*{-10mm}
	\caption*{FIG.S1: EDX spectra with atomic percentage of Cs$_2$TeCl$_6$ at two different positions in ambient conditions}
	\includegraphics[scale=0.75]{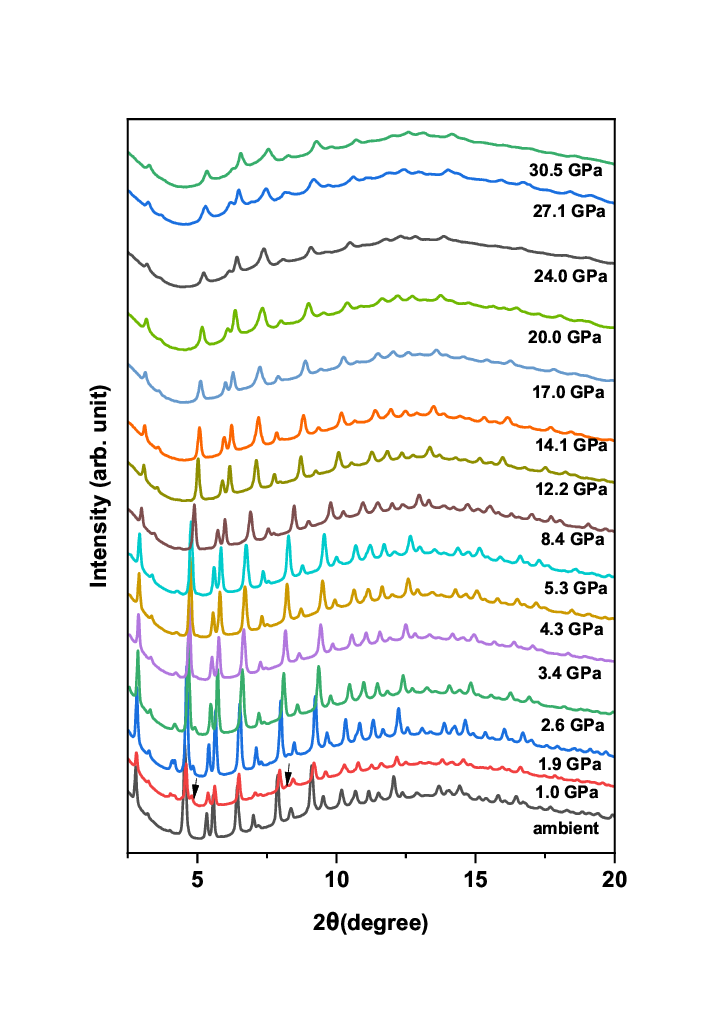}
	\vspace{-10mm}
	\caption*{FIG.S2: XRD spectra of Cs$_2$TeCl$_6$ at some selected pressure points}
\end{figure}

\begin{figure*}[ht]
	
	\centering
	\includegraphics[scale=0.8]{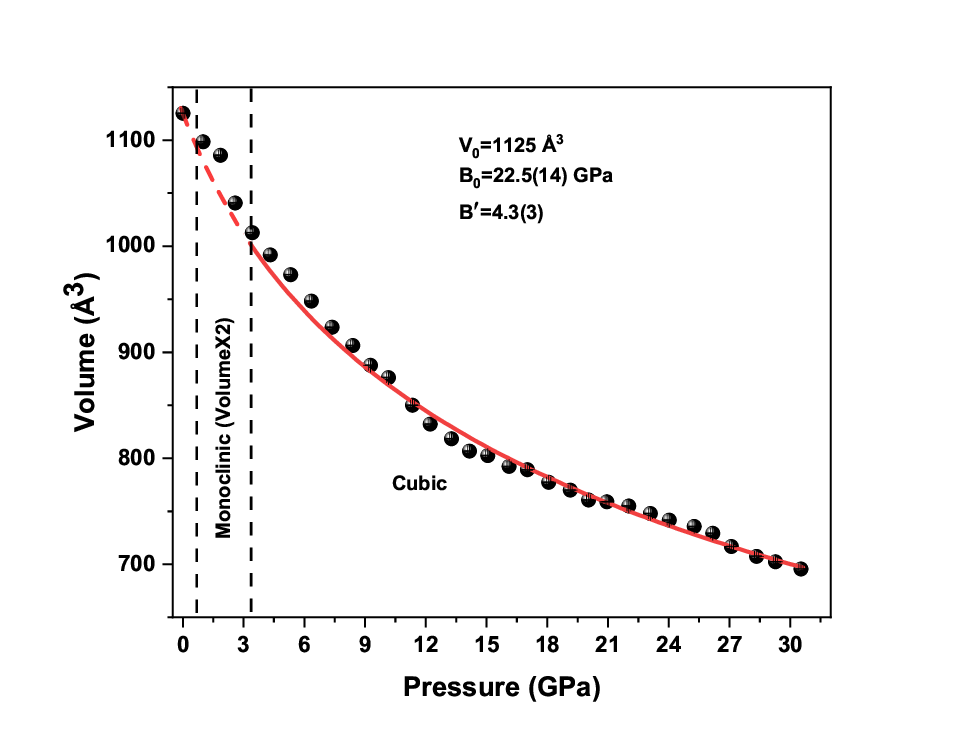}
	\caption*{FIG.S3: Pressure evolution of the unit cell volume of Cs$_2$TeCl$_6$ using a single 3rd order Birch–Murnaghan EoS.}
\end{figure*}

\begin{figure}[ht]
	\centering
	\includegraphics[scale=0.8]{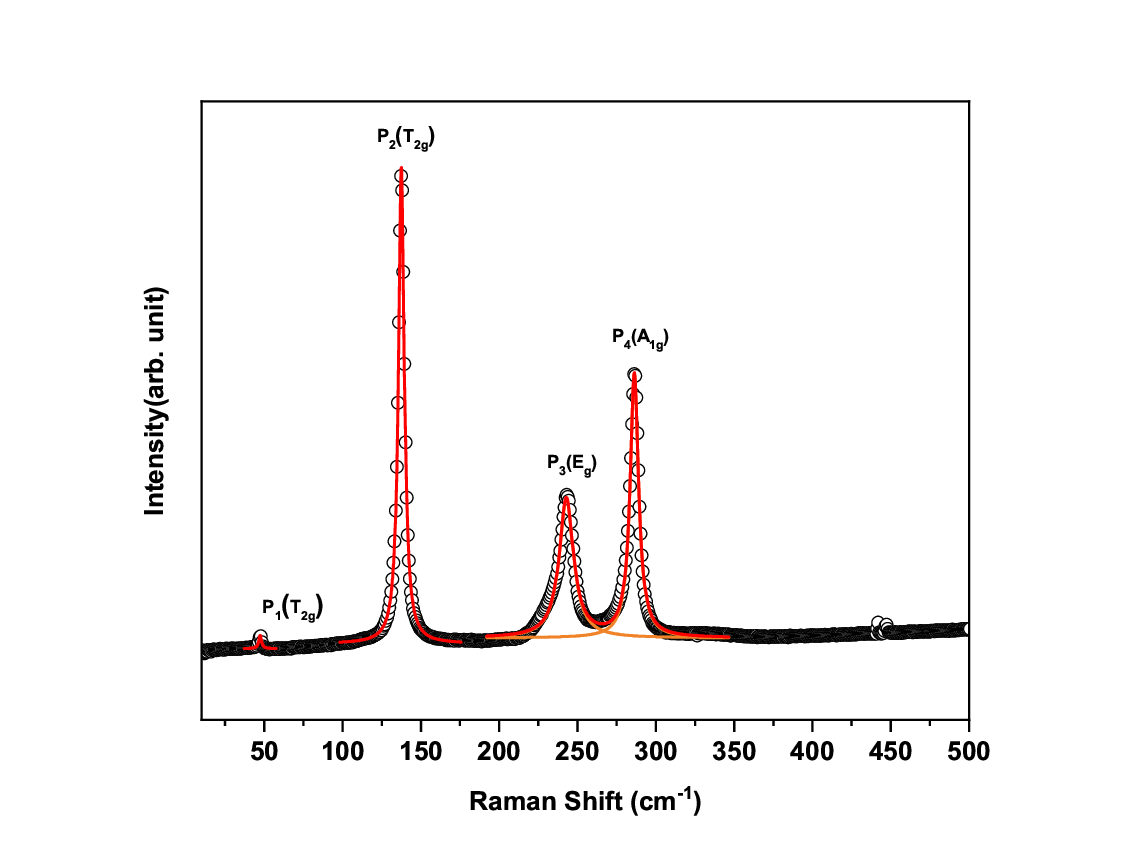}
	\caption*{FIG.S4: Raman spectrum of Cs$_2$TeCl$_6$ in ambient condition. The black circles represent the experimental data and the red line shows a fit of experimental data points to Lorentzian functions.}
\end{figure}

\begin{figure}[ht]
	\centering
	\includegraphics[scale=0.8]{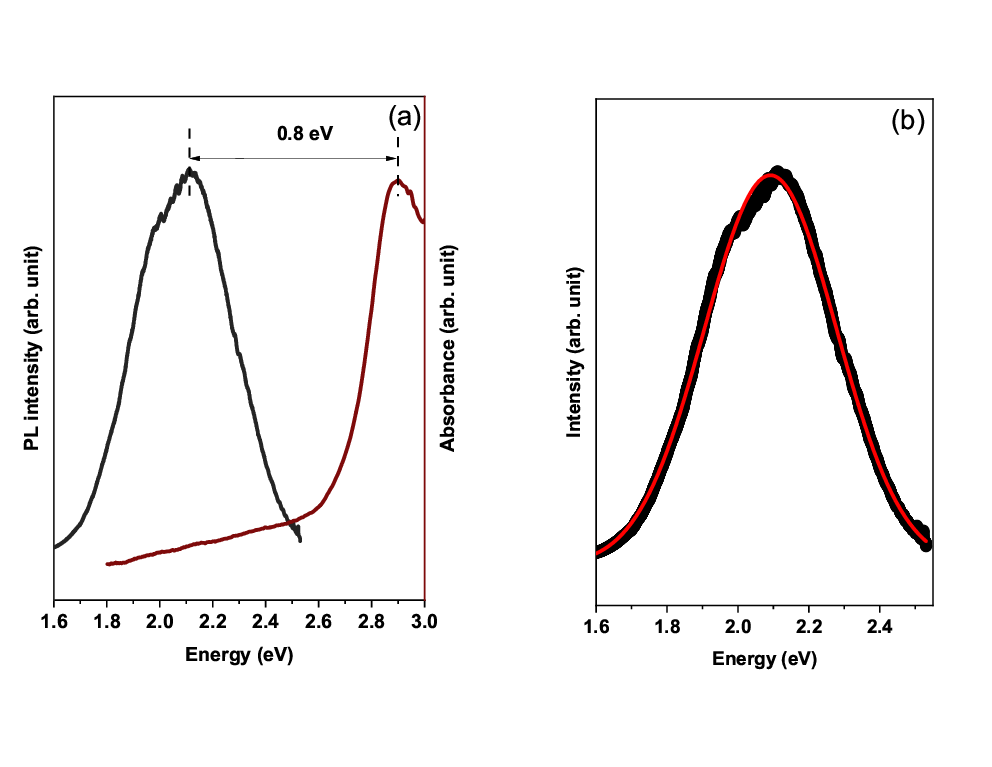}
	\caption*{FIG.S5: (a).Stokes shift of Cs$_2$TeCl$_6$ in ambient condition.(b) Ambient PL spectrum of Cs$_2$TeCl$_6$.The black balls represent the experimental data and the red line shows a fit of the experimental data points to Gaussian function}
\end{figure}

\begin{figure}[ht]
	\centering
	\includegraphics[scale=0.8]{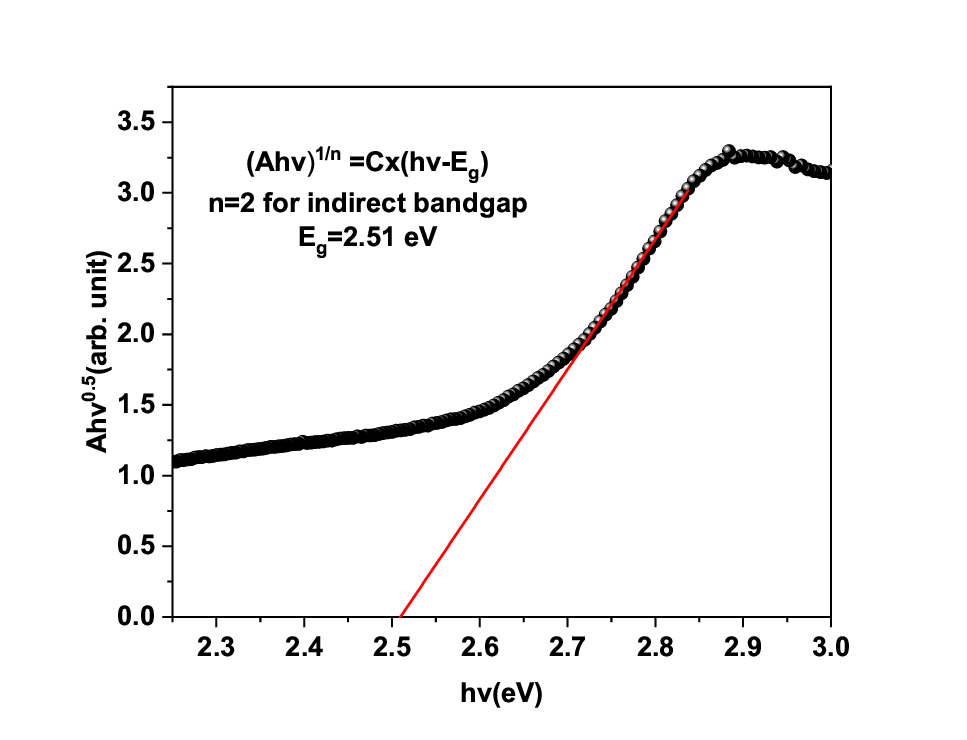}
	\caption*{FIG.S6: Determination of the optical bandgap of Cs$_2$TeCl$_6$ from the absorption spectrum in ambient condition.}
\end{figure}

\end{document}